\documentclass[twoside,11pt]{article}

\usepackage[svgnames]{xcolor}
\usepackage[colorlinks=true, linkcolor=Maroon, citecolor=MidnightBlue]{hyperref}

\usepackage{hamid}
\usepackage{booktabs} % For formal tables

\usepackage[numbers]{natbib}

\usepackage{times}  %Required
\usepackage{helvet}  %Required
\usepackage{courier}  %Required
\usepackage{url}  %Required
\usepackage{graphicx}  %Required

\usepackage{algorithm}
\usepackage{algorithmic}

\usepackage[T1]{fontenc}    % use 8-bit T1 fonts
\usepackage{amsfonts}       % blackboard math symbols
\usepackage{nicefrac}       % compact symbols for 1/2, etc.
\usepackage{amsmath}
\usepackage{rotating}
\usepackage{subcaption}
\usepackage{color}
\usepackage{flushend}

\begin{document}
% Title portion. Note the short title for running heads 
\title{Alternating Linear Bandits\\ for Online Matrix-Factorization Recommendation}  

\author{\name Hamid Dadkhahi \\
       \addr College of Information and Computer Sciences\\
       University of Massachusetts Amherst \\
       \AND
       \name Sahand Negahban \\
       \addr Department of Statistics and Data Science\\
       Yale University}

\maketitle

\begin{abstract}

We consider the problem of online collaborative filtering in the online setting, where items are recommended to the users over time. At each time step, the user (selected by the environment) consumes an item (selected by the agent) and provides a rating of the selected item.
In this paper, we propose a novel algorithm for online matrix factorization recommendation that combines linear bandits and alternating least squares. In this formulation, the bandit feedback is equal to the difference between the ratings of the best and selected items. 
We evaluate the performance of the proposed algorithm over time using both cumulative regret and average cumulative NDCG. Simulation results over three synthetic datasets as well as three real-world datasets for online collaborative filtering indicate the superior performance of the proposed algorithm over two state-of-the-art online algorithms.

\end{abstract}

\section{Introduction}

% paragraph 1: collaborative filtering and matrix factorization; intro to general MF notation

% In collaborative filtering, given a set of ratings for different users, the aim is to predict the ratings for the unrated items for each user. 

Matrix factorization casts the foundation of some of the most successful realizations of collaborative filtering \cite{Koren:2009}. The most convenient data input to a collaborative filtering system is in terms of explicit feedback (\emph{ratings}), which comprises a sparse matrix $Y \in \mathbb{R}^{n \times m}$.
In its basic formulation, matrix factorization decomposes the aforementioned matrix $Y$ into the product of two smaller matrices $A \in \mathbb{R}^{n \times k}$ and $B \in \mathbb{R}^{m \times k}$, i.e.
$Y = A \cdot B^{\top}$.
Since its inception, a plethora of variations of matrix factorization based methods have been proposed in the literature. 
Despite the success of such methods, there has been little development to adapt their employment to the online setting, where items are recommended to users over time. This is specifically a difficult problem due to the explore-exploit dilemma, where the system needs to explore new items while quickly adapting to the user feedback in order to match users' interests.

% paragraph 2: problems with offline matrix factorization; intro to online matrix factorization/collaborative filtering

%This setting was introduced in the previous work \cite{PTS} based on Thompson Sampling, \cite{NMF-Bandit} based on epsilon greedy.

In recent years, a few efforts have been made to devise online matrix factorization algorithms for collaborative filtering via borrowing ideas from the bandit literature. In particular, \cite{NMFBandit} propose an algorithm based on an $\epsilon$-greedy approach, while \cite{PTS}  offers an approach using Thompson Sampling.
In the former case, bandit algorithms based on $\epsilon$-greedy are known to be sub-optimal. In the latter case, PTS does not have a closed form solution and requires resorting to approximate sampling methods which could potentially introduce additional errors to the matrix factorization. 

% paragraph 3: our algorithm

In this work, we address these issues by formulating the online matrix factorization problem as an upper confidence bound (UCB) based bandit algorithm. 
To the best of our knowledge, this is the first effort to an online matrix factorization algorithm based on UCB-type bandit framework when no additional side information about the users or the items is available. We model the problem as a linear bandit, where user and item feature vectors correspond to bandit parameter and action vectors, respectively. The proposed algorithm, referred to as alternating linear bandits algorithm, alternates between learning the user and item feature vectors, in a fashion reminiscent of the alternating least squares.
Numerical experiments over three real world datasets indicate that the proposed algorithm outperforms the state-of-the-art online algorithms in terms of both cumulative regret as well as average cumulative NDCG.

\section{Related Work}

The majority of the literature on collaborative filtering addresses the offline setting. 
In addition, the literature on active collaborative filtering require additional prior knowledge such as user types (e.g. \cite{Harpale:2008}) and item features (e.g. \cite{Zhao:2013}), which are in many scenarios unavailable in advance. \cite{Deshpande2012} considers collaborative filtering as a linear bandit problem, but does not use any collaboration beyond a pre-processing step in which offline collaborative filtering (specifically matrix completion) is solved to compute feature vectors for items.
In the online setting, \cite{Dabeer2013} models the online collaborative filtering by a Markov decision process, but assumes that the numbers of user types and item types are known. 
%More recent efforts have employed algorithms from the bandit literature to model the exploration-exploitation trade-off required in the online setting. 

FactorUCB \cite{Wang2017} is a factorization-based bandit algorithm where an upper confidence bound based item selection strategy is used for online recommendation. The problem setup considered in \cite{Wang2017} is different from this work in that they leverage additional side information in form of contextual features as well as dependency among users in order to improve the convergence of the algorithm. In the former case, they introduce an observable contextual feature factor into the estimation process, which extends the item feature factor and its true value is provided to the algorithm.
In the latter case, they assume that the observed reward from each user is determined by a mixture of neighboring users, as given by a weight matrix which encodes the affinity relation among users. In addition to incorporation of such side information in the framework, FactorUCB is different from our proposed algorithm in that Alternating Least Squares (ALS) is employed for updating both user and item feature factors. Furthermore, they define confidence sets for both user and item feature factors; upper confidence bounds for both user and item feature factors are then incorporated into the optimization objective for the item selection strategy. On the contrary, we only define confidence sets for the user feature vectors and use the joint OFUL optimization for both selecting the item and finding new estimates for the user feature factors. Least squares is then employed to find a new estimate for the item feature vectors.

Valko et al. \cite{Valko:2014} proposed a bandit algorithm for content-based recommendations. In this approach, the features of the items are extracted from a similarity graph over the items, which is known in advance. Gentile et al. \cite{Gentile:2014} propose a bandit algorithm for recommendations that clusters users in an online fashion based on the similarity of their preferences. The preferences are learned by regressing the ratings of the items from their features. The features of the items are the input of the learning algorithm and they only learn $A$. Maillard et al. \cite{Maillard:2014} study a bandit problem where the arms are partitioned into unknown clusters unlike our work which is more general.
Collaborative-Greedy (CG) algorithm of \cite{Chen} uses an $\epsilon$-greedy algorithm with three types of stages, namely joint exploration (for clustering of users), random exploration, and exploitation. However, \cite{Chen} enforces two major restrictions in that the feedback is limited to the binary case (likable versus unlikable items) and that the previously recommended items cannot be repeated for a given user in the future. 

In particular, our work is closely related to the Non-negative Matrix Factorization Bandit (NMF-Bandit) algorithm \cite{NMFBandit} and the Particle Thompson Sampling (PTS) \cite{PTS} in terms of the problem setup. Sen et al. \cite{NMFBandit} model the collaborative filtering problem as a contextual bandit with a latent low-dimensional confounder.  They combine an $\epsilon$-greedy algorithm with non-negative matrix factorization in order to achieve a balance between learning this low-dimensional structure and selecting the best arm to minimize the regret. More specifically, at each time step $t$, the algorithm explores with a given probability, where an item is selected at random in a user (context)-dependent or independent manner, which is aimed at facilitating the estimation of the user or item feature factors, respectively. Alternatively, the algorithm exploits, where the best item is selected according to the new estimates obtained for the user and item feature matrices and hence the ratings matrix.
Kawale et al. \cite{PTS} propose a Thompson Sampling framework to model online Bayesian matrix factorization. This formulation does not have a closed form solution and they use Rao-Blackwellized particle filtering in order to design a computationally efficient solution for the problem of huge parameter space.

\section{Alternating Linear Bandits}
\label{sec:algorithm}

\textbf{Notation}. Sets  are shown with calligraphic letters. Matrices are indicated with uppercase letters; vectors and scalars are indicated with lowercase letters. Let $[n] = \{1,2,\ldots,n\}$. For a matrix $A \in \mathbb{R}^{n \times m}$, $A_{i, j}$ indicates the entry in row $i$ and column $j$, and $A_i$ designates the $i$-th row of $A$. Also, for an index set $\mathcal{I} \subset [n]$, $A_{\mathcal{I}}$ indicates a sub-matrix of $A$ with rows in the index set $\mathcal{I}$. We use $\|x\|_p$ to denote the $\ell_p$ norm of a vector $x \in \mathbb{R}^k$, and denote the inner product by $\langle \cdot, \cdot \rangle$. For a positive definite matrix $A \in \mathbb{R}^{k \times k}$, the weighted $\ell_2$ norm is defined by $\|x\|_A = \sqrt{x^{\top} A x}$.

\textbf{The Learning Model}. We consider an online collaborative filtering problem with $n$ users and $m$ items. At each time step $t$, a given user $i_t$ is offered an item $j_t$; the user in turn gives a feedback via a rating $y_t = Y_{i_t, j_t} + \eta_t$, where $\eta_t$ is an additive noise term. We assume that the noise sequence $\{\eta_t\}$ is a sub-Gaussian process with zero mean and variance of $\sigma^2$ conditioned on previous actions and noise terms. We characterize the ratings matrix $Y \in  \mathbb{R}^{n \times m}$ as a product of rank-$k$ user $A^* \in \mathbb{R}^{n \times k}$ and item $B^* \in \mathbb{R}^{m \times k}$ feature factors, i.e. $Y = A^* \cdot {B^*}^{\top}$. We model this problem as a repeated game in a bandit setting where the environment chooses row $i_t$ of $Y$ and the learning agent chooses column $j_t$. The reward (rating) $y_t $ is revealed and the goal of the learning agent is to minimize the cumulative regret with respect to the optimal solution, the highest entry in row $i_t$ of matrix $Y$. 
The cumulative regret of the problem until time $T$ is given by:
\begin{equation}
\textrm{Regret}(T)  = \sum_{t=1}^T \left(  Y_{i_t j^*_t} - y_t    \right) = \sum_{t=1}^T \left( \max_{j \in [m]} A^*_{i_t} {B^*_{j}}^{\top} - y_t  \right)
\label{eq:regret}
\end{equation}
where $j^*_t$ is the best item for user $i_t$, i.e. $j^*_t = \arg\max_{j \in [m]}  Y_{i_t, j}$. 

\textbf{The Proposed Algorithm}. We model the problem as a linear bandit problem, where user and item feature vectors correspond to bandit parameter and action vectors, respectively. A natural way to solve this linear bandit problem is via the principle of optimism in the face of uncertainty linear bandit algorithm (OFUL)  \cite{OFUL}. 
The main idea behind the OFUL algorithm is to maintain a confidence set $\mathcal{C}_t \subset \mathbb{R}^k$ for the bandit parameter vector over time. In addition, the decision set $\mathcal{D}_t$ is defined as the set of possible actions at time $t$. At any given time $t$, the confidence set $\mathcal{C}_t$ is constructed from the history of actions and observations such that the parameter vector lies in $\mathcal{C}_t$ with high probability. The algorithm then chooses an optimistic estimate for the pair of action and parameter vector which jointly maximizes the reward.

We adopt OFUL in our problem in the following way. The decision set  $\mathcal{D}_t$ corresponds to the set of rows of matrix $B$: 
\begin{equation}
  \mathcal{D}_t = \{ B_j : \forall j \in [m] \}
  \label{decision_set}
\end{equation}
where $B$ is our current estimate for the true item feature matrix $B^*$. Intuitively, selection of each action (i.e. element of the decision set) corresponds with the selection of the relevant item in the recommender system.
Let $X$ be a matrix where row $X_t$ represents the feature vector (action) for the item selected at time $t$.
%Let the sequence $\{X_t\}$ represent the feature vector (actions) for the items selected over time. We then generate matrix $X$ by stacking the sequence elements $X_t$ as its rows. 
The confidence set $\mathcal{C}_t$ is constructed as a high-dimensional ellipsoid, centered at 
\begin{equation}
  \mu = (X_{\mathcal{I}}^{\top} \cdot X_{\mathcal{I}} + \lambda_1 I_k )^{-1} X_{\mathcal{I}}^{\top} y_{\mathcal{I}}
\end{equation}
where $I_k$ is the identity matrix of size $k \times k$ and $\lambda_1 > 0$ is a regularization parameter. Note that in computing the ellipsoid center $\mu$, we have only used the previous actions $ X_{\mathcal{I}}$ and observations $y_{\mathcal{I}}$ for user $i_t$, indexed by $\mathcal{I} = \{ \ell  < t : i_{\ell} =  i_t\}$. As a result, the confidence set is defined as 
\begin{align}
  \label{confidence_set}
  \mathcal{C}_t &= \{q \in \mathbb{R}^k : \|\mu - q\|_{V}\leq c\}  \\
   c &= \sigma \sqrt{2 \log\left ( \tfrac{\det(V)^{1/2} \det(\lambda_1 I_k)^{-1/2}}{\delta} \right )} + \lambda_1^{1/2} s  \nonumber
\end{align}
where $s = \max_{i \in [n]} \|A_i\|_2$ is the maximum $\ell_2$ norm of the rows of $A$, $V$ is  a positive semi-definite matrix defined as $V = \lambda_1 I_k + \sum_{\ell \in \mathcal{I}} X_{\ell}^{\top} \cdot X_{\ell}$, and $1-\delta$ is the probability with which the parameter vector belongs to the confidence set $\mathcal{C}_t$.
% singular values of $V^{1/2}$ are square roots of those of $V$.

Having defined the decision and confidence sets in \eqref{decision_set} and \eqref{confidence_set}, we formulate the joint OFUL optimization as follows:
\begin{equation} 
(j_t, A_{i_t}) = \arg\max_{(j, q) \in \mathcal{D}_t \times \mathcal{C}_t} \langle q, B_j \rangle
\label{oful}
\end{equation} 
whose solution is given by
\begin{align}
j_t &= \arg\max_{j \in [m]} \mu \cdot B_j^{\top} + c \|V^{-1/2} \cdot B_j^{\top} \|_2 \\
A_{i_t} &= \arg\max_{q \in C_t} q \cdot B_{j_t}^{\top} = \mu + c \tfrac{V^{-1} \cdot B_{j_t}^{\top}}{ \| V^{-1/2} \cdot B_{j_t}^{\top} \|_2}.
\end{align}
Having selected the item $j_t$ and obtained a new estimate for the user feature vector $A_{i_t}$,  the item $j_t$ is offered to the user $i_t$, who in turn gives a rating $y_t = Y_{i_t, j_t} + \eta_t$ for the item $j_t$.

Next, we derive a new estimate for the feature vector of the selected item $j_t$ via regularized least squares:
\begin{eqnarray}
  B_{j_t} = (Z_{\mathcal{J}}^{\top} \cdot Z_{\mathcal{J}} + \lambda_2 I_k )^{-1} Z_{\mathcal{J}}^{\top} y_{\mathcal{J}}
 \label{least_squares}
\end{eqnarray}
where $Z$ represents a matrix whose rows are the sequence of user feature vectors over time, and $Z_{\mathcal{J}}$ is a sub-matrix whose rows are the elements $Z_{\ell}: \forall \ell \in \mathcal{J}$. In computing the estimate for $B_{j_t}$, we have only utilized the feature vectors of users who have rated the item $j_t$, indexed by $\mathcal{J} = \{ \ell  \leq t: j_{\ell} =  j_t\}$. In addition, $\lambda_2 > 0$ is a regularization parameter.

Alternating between OFUL steps given by Equation \eqref{oful} and least squares steps given by Equation \eqref{least_squares} forms our algorithm, which we refer to as \textit{Alternating Linear Bandits} (ALB). A summary of ALB is given in Algorithm \ref{algo:ALB}.
Note that we start the algorithm with random initializations for both user and item feature matrices $A$ and $B$. In addition, as we obtain improved estimates at any time step $t$ for the user and item feature vectors $A_{i_t}$ and $B_{j_t}$, we update the rows of matrices $Z$ and $X$ corresponding to the user $i_t$ and item $j_t$, respectively (See lines $16$ and $18$ of Algorithm \ref{algo:ALB}).

\begin{algorithm}[!t]
\caption{Alternating Linear Bandits}
\begin{algorithmic}[1]
\STATE \textbf{Inputs:} horizon $T$, user sequence $\{i_t\}$, true reward matrix $Y$, number of users $n$, number of items $m$, rank of the model $k$, regularization parameters $\lambda_1$ and $\lambda_2$, noise parameter $\sigma$, probability of error $\delta$
\STATE \textbf{Outputs:} Regret sequence $\{r_t\}$
\STATE $\forall i, j: A_{ij} \sim \mathcal{N}(0, \sigma_1), B_{ij} \sim \mathcal{N}(0, \sigma_2)$
\FOR {$t = 1:T $}
\STATE $\mathcal{I} \gets \{ \ell  < t : i_{\ell} =  i_t\}$
\STATE $\mu \gets (X_{\mathcal{I}}^{\top} \cdot X_{\mathcal{I}} + \lambda_1 I_k )^{-1} X_{\mathcal{I}}^{\top} y_{\mathcal{I}}$
\STATE $V \gets \lambda_1 I_k + \sum_{\ell \in \mathcal{I}} X_{\ell}^{\top} \cdot X_{\ell}$
\STATE $c \gets \sigma \sqrt{2 \log\left ( \tfrac{\det(V)^{1/2} \det(\lambda_1 I_k)^{-1/2}}{\delta} \right )} + \lambda_1^{1/2} s$
\STATE $\mathcal{C}_t \gets \{q \in \mathbb{R}^k : \|\mu - q\|_{V}\leq c\}$
\STATE $\mathcal{D}_t \gets \{ B_j : \forall j \in [m] \}$ 
\STATE $ (j_t, A_{i_t}) \gets \arg\max_{(j, q) \in \mathcal{D}_t \times \mathcal{C}_t} \langle q, B_j \rangle$ \hfill \COMMENT{OFUL Step}
\STATE $\textrm{Play item}~j_t~\textrm{for user}~i_t$
\STATE $\textrm{Observe reward}~y_t = Y_{i_t, j_t} + \eta_t$
\STATE $r_t \gets Y_{i_t, j^*_t} - y_t$ 
\STATE $\mathcal{J} \gets \{ \ell  \leq t: j_{\ell} =  j_t\}$
\STATE $\forall \ell \in \mathcal{I}: Z_{\ell} \gets A_{i_t}$
\STATE $B_{j_t} \gets (Z_{\mathcal{J}}^{\top} \cdot Z_{\mathcal{J}} + \lambda_2 I_k )^{-1} Z_{\mathcal{J}}^{\top} y_{\mathcal{J}}$ \hfill \COMMENT{LS Step}
\STATE $\forall \ell \in \mathcal{J}: X_{\ell} \gets B_{j_t}$
%\STATE $\texttt{Update}~\mathcal{D}_t~\texttt{and}~\mathcal{C}_{t}$
\ENDFOR
\RETURN $\{r_t\}$
\end{algorithmic}
\label{algo:ALB}
\end{algorithm}

\textbf{Computational Complexity}. The computational complexity of the algorithm at each time step $t$ is dictated by lines $6$ and $17$ of the algorithm, and is in $\mathcal{O} \big ( (m' + n') \cdot k^2 \allowbreak + k^3 \big )$, where $m'$ and $n'$ are the number of recommendations (of any item) made to user $i_t$ and the number of times item $j_t$ has been consumed (by any user), respectively, both until time $t$. The first term linearly increases with time, but this can be mitigated by the rarely switching version of the OFUL algorithm \cite{OFUL}. The dependency on $k^3$ is due to matrix inversion, but this is not a concern since the model rank $k$ is typically small \footnote{For the experiments in this paper, the running time of ALB is approximately the same as that of PTS (in the order of a few minutes for $T=25000$), both of which are three to four times faster than NMF-Bandit.}.

\section{Experiments and Results}

In this section, we evaluate our algorithm on three synthetic datasets as well as three real world datasets and compare its performance to two state-of-the-art online\footnote{For a discussion on the performance of online methods in comparison to offline methods, see \cite{PTS}.
} algorithms: PTS \cite{PTS} and NMF-Bandit \cite{NMFBandit}. 
%We evaluate the performance of different algorithms using cumulative regret, as measured via the formula given in Equation \ref{eq:regret}.

The selected real world datasets are Book Crossing, MovieLens, and Jester. We selected these real world datasets since they have a diverse density of available ratings, where Book Crossing, MovieLens, and Jester are sparse, moderately sparse, and dense, respectively; the ratio of the number of ratings available divided by the number of all possible ratings (i.e. the number of users times the number of items) is $0.002$, $0.058$, and $0.373$ for the three datasets, respectively. In addition, the first two datasets have been used in both \cite{PTS} and \cite{NMFBandit}; hence, we already have an initial estimate for the optimal hyper-parameter values as given in the respective references to ensure a fair comparison.

We evaluate the performance of different algorithms using two metrics: cumulative regret and average cumulative NDCG@k. In the case of the former metric, we use the cumulative regret formula given in Equation \eqref{eq:regret}.
In the case of the latter metric, we measure average cumulative NDCG@k (with $k = 5$) against time, where average cumulative NDCG@k at time $t$ is the cumulative NDCG@k up to time $t$, divided by $t$.
For NDCG computations, we shift the minimum ratings to zero, since relevance scores need to be non-negative in order for the NDCG score to be between zero and one (and hence interpretable). This only applies to Jester dataset, as the other two datasets have non-negative ratings.

We set the matrix factorization rank $k = 5$ for all the algorithms in all the experiments, unless specified otherwise. Experimentally, as we will see at the end of this section, tuning the rank $k$ for different algorithms would only marginally improve the results. We set the parameters $\sigma_1$ and $\sigma_2$ used for the initialization (used in ALB) of user and item feature matrices $A$ and $B$ to $1$ in all the experiments. We also set the probability of error $\delta = 0.01$ in all the experiments. Furthermore, we set the scaling factor $s$ to $1$. Finally, we set the prior hyper-parameters of PTS, i.e. $\sigma_1$ and $\sigma_2$, to $1$ in all the experiments, as suggested in \cite{PTS}.

In order to have a fair comparison of the performance of different algorithms, we find the optimal hyper-parameters of each algorithm via grid search for any given dataset, and only report the best (minimum) cumulative regret. For each algorithm, we have two hyper-parameters to be tuned as follows. For ALB, we tune the regularization parameter $\lambda = \lambda_1 = \lambda_2$ \footnote{Note that the optimal regularization parameters of ALB are not necessarily equal. We made this choice for simplicity. Further tuning of the regularization parameters would only improve the results of the proposed algorithm further.} as well as the noise parameter $\sigma$. 
%The grid search for the two hyper-parameters are carried out over the sets $\{0.001, 0.01, 0.1, 1\}$ and $[0.1, 1]$ (with step size of $0.1$), respectively. 
In the case of PTS, we tune the number of particles as well as the noise parameter $\sigma$. 
%The grid search interval for the former and latter hyper-parameters are $\{10, 20, 30, 40, 50\}$ and $[0.1, 1]$ (with step size of $0.1$), respectively. 
Finally, for the NMF-Bandit algorithm, we tune the hyper-parameters $\theta$ (the exploration parameter) and $m'$ (size of the exploration set for each user, in their notation).
%, where the grid search is performed over the sets $\{1, 2, 3, 4, 5\}$ and $\{5, 10, 15, 20, 25, 30\}$, respectively.

\begin{figure*}[t]
\center
\begin{tabular}{ccc}
\includegraphics[width=1.9in]{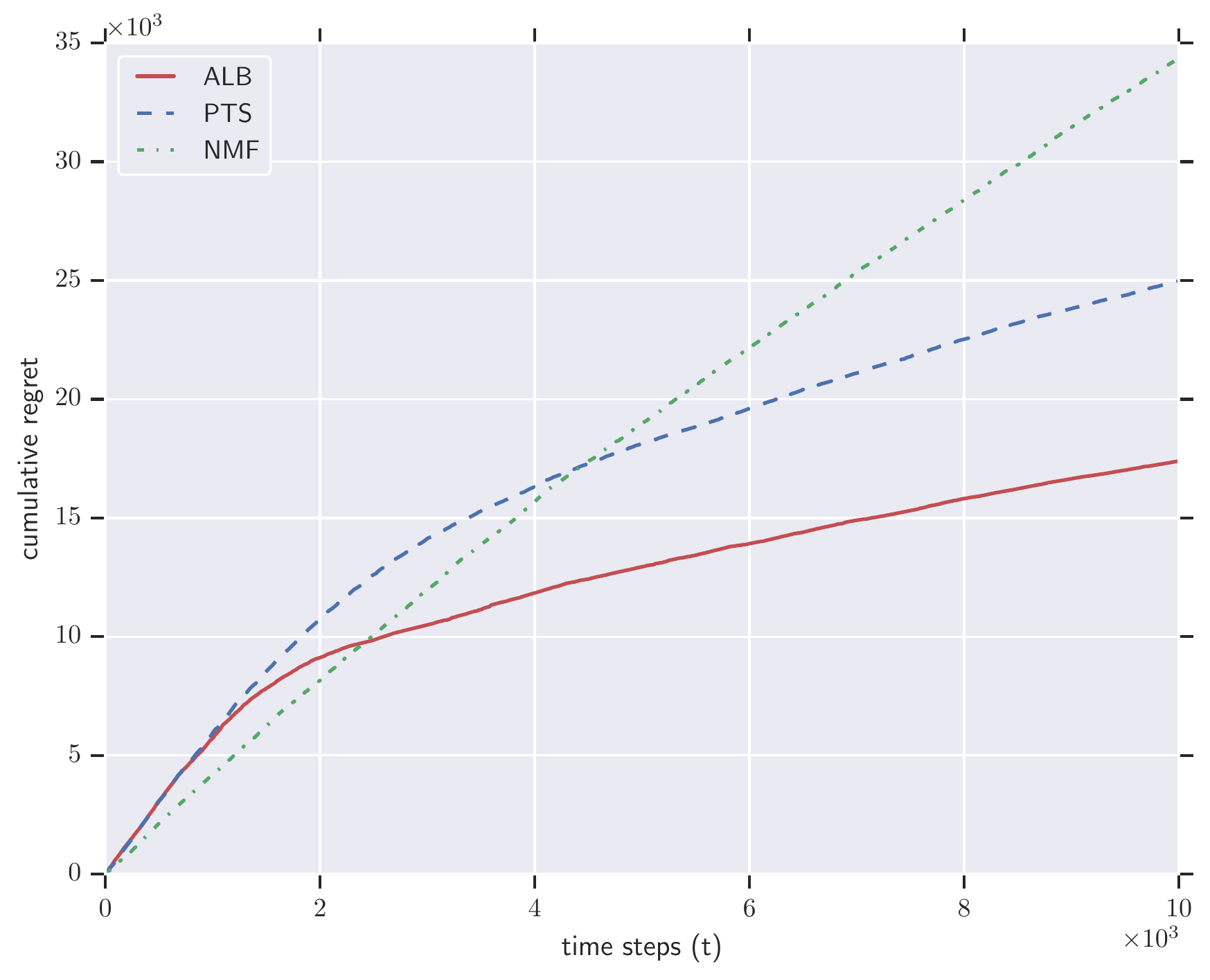} 
&
\includegraphics[width=1.9in]{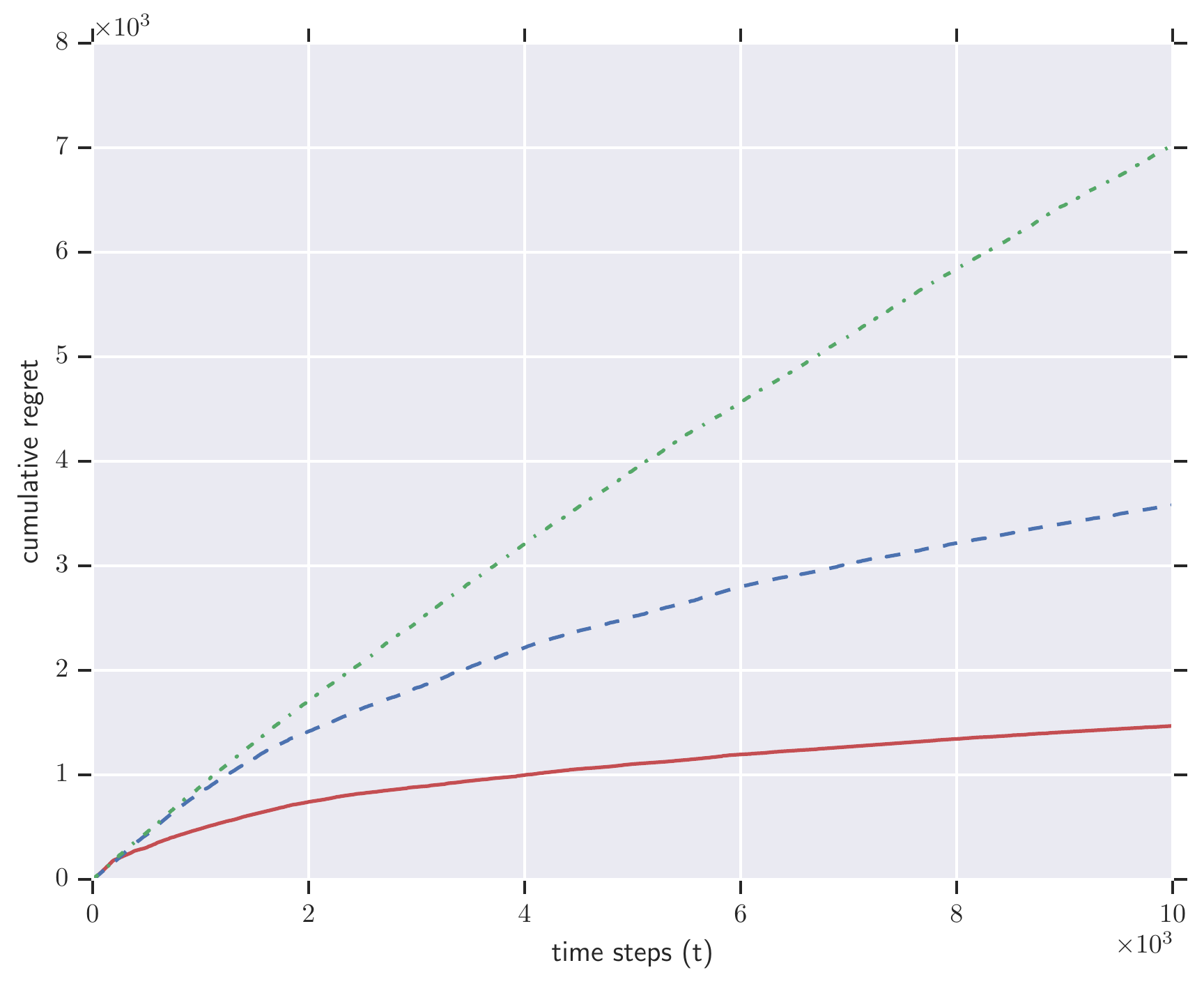}
&
\includegraphics[width=1.9in]{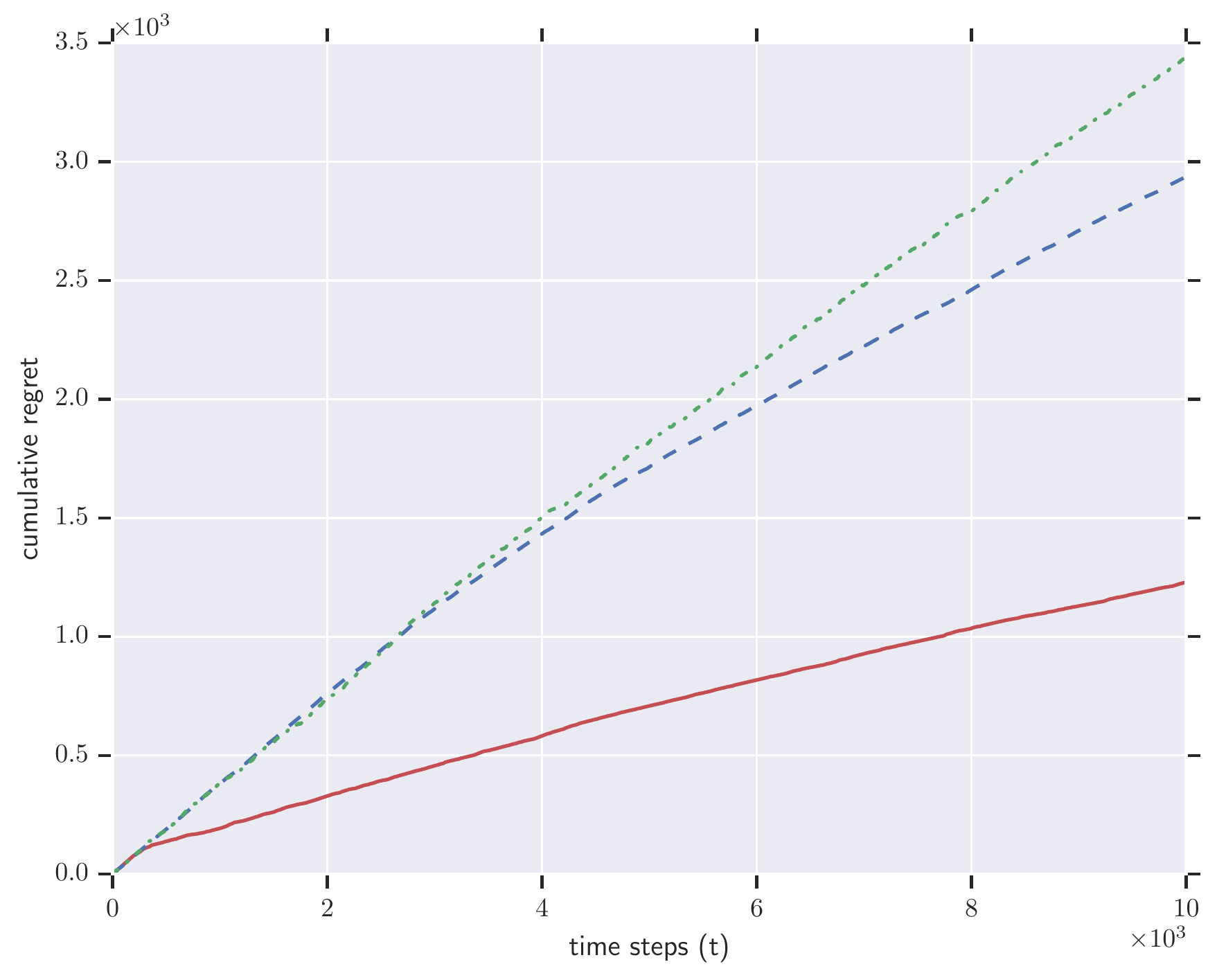} 
\end{tabular}
\caption{\small \label{fig:synthetic}
Evaluation of different algorithms in terms of cumulative regret for synthetic datasets: Gaussian (left), Uniform (middle), and Bernoulli (right).}
\end{figure*}

\begin{figure*}[t]
\center
\begin{tabular}{ccc}
\includegraphics[width=1.9in]{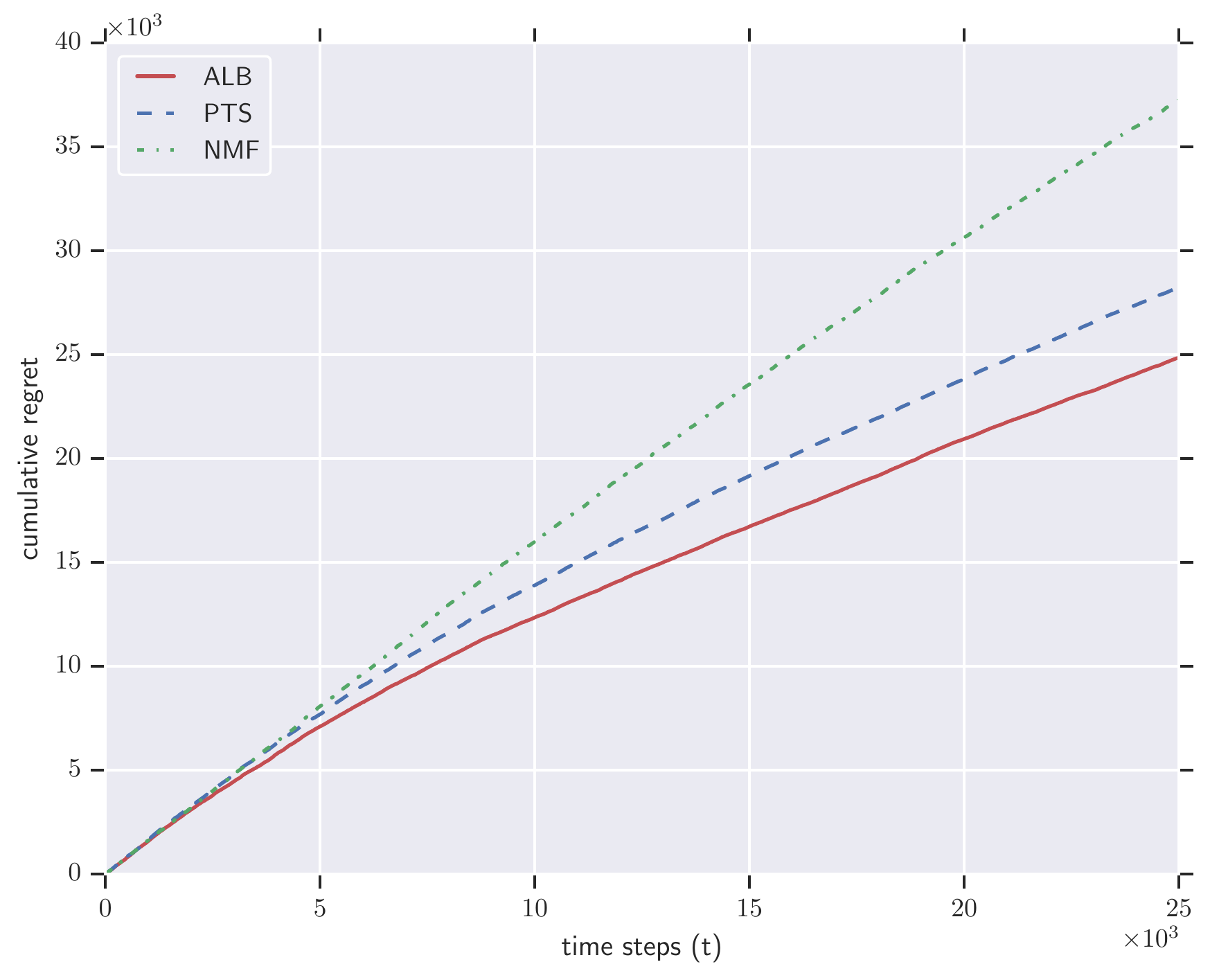} 
&
\includegraphics[width=1.9in]{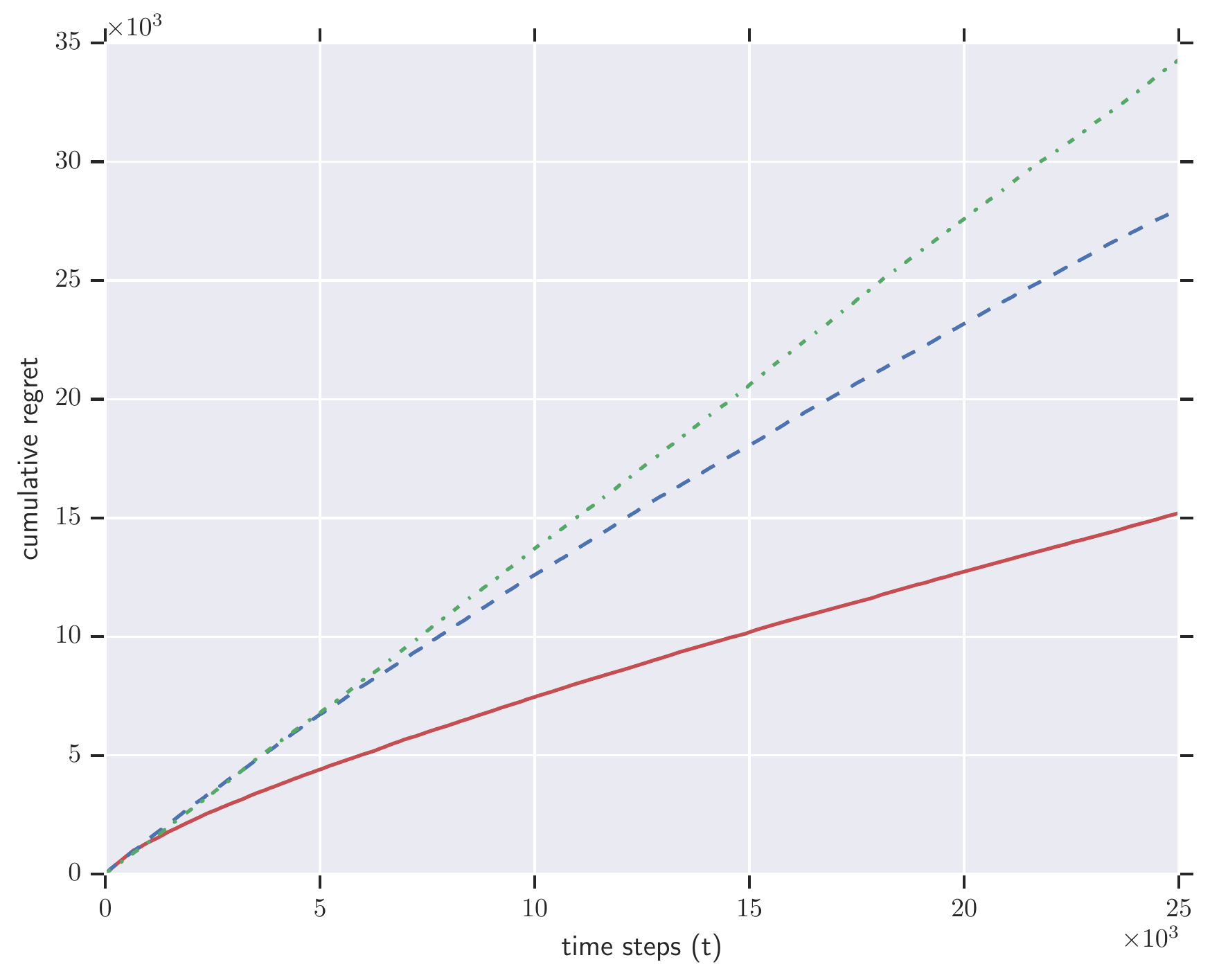}
&
\includegraphics[width=1.9in]{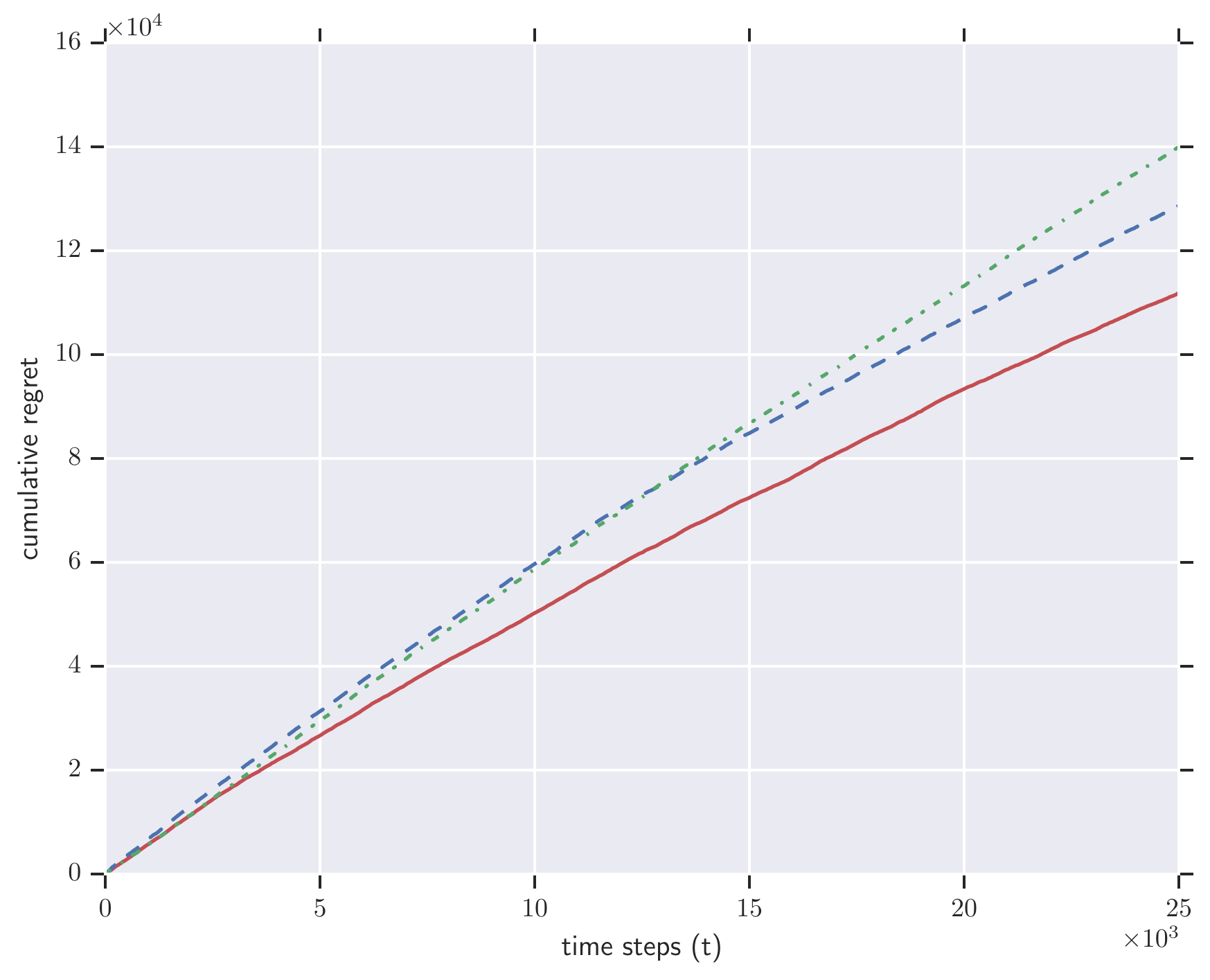} 
\end{tabular}
\caption{\small \label{fig:real}
Evaluation of different algorithms in terms of cumulative regret for real-world datasets: Book Crossing (left), MovieLens (middle), and Jester (right).}
\end{figure*}

\begin{figure*}[t]
\center
\begin{tabular}{ccc}
\includegraphics[width=1.9in]{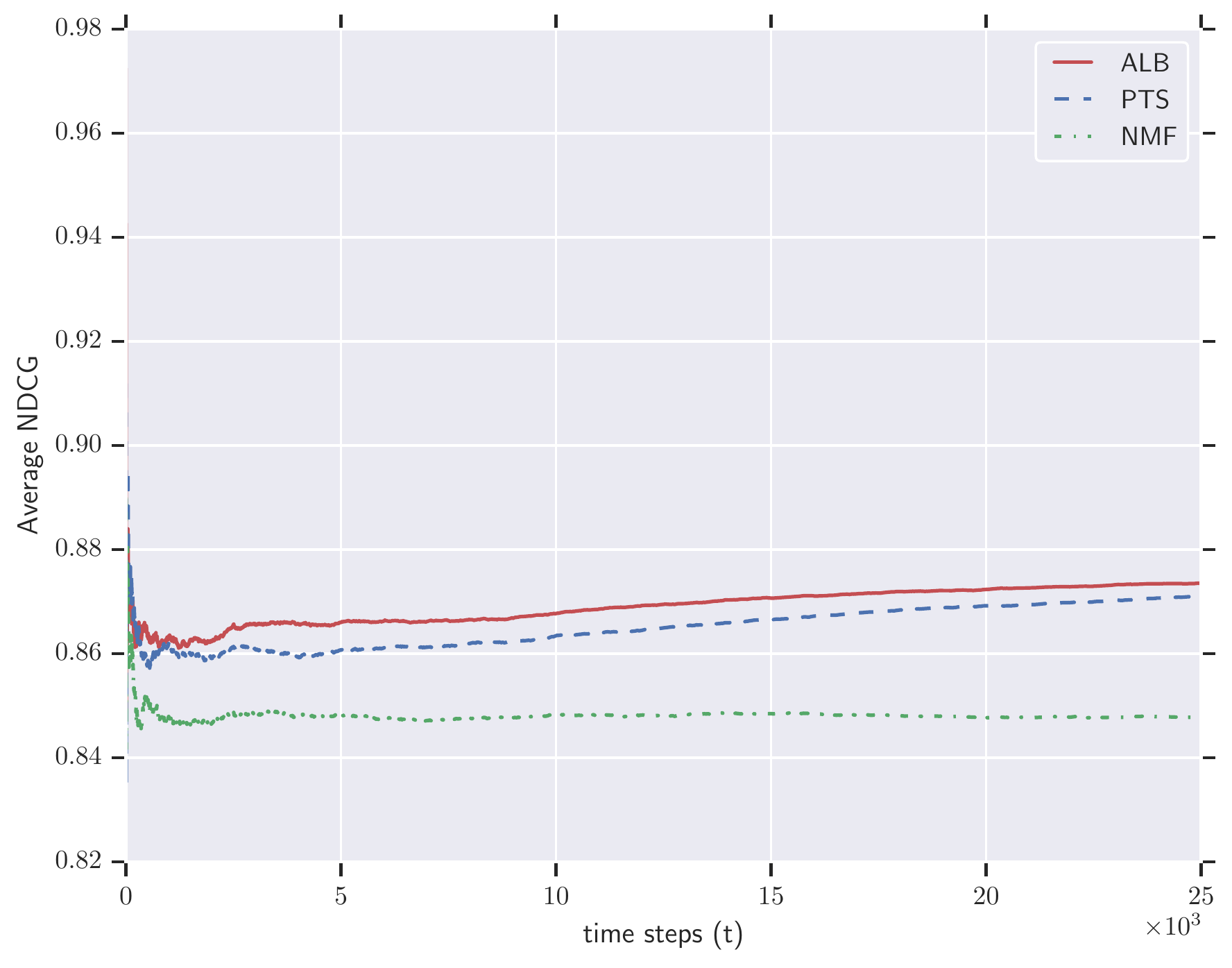} 
&
\includegraphics[width=1.9in]{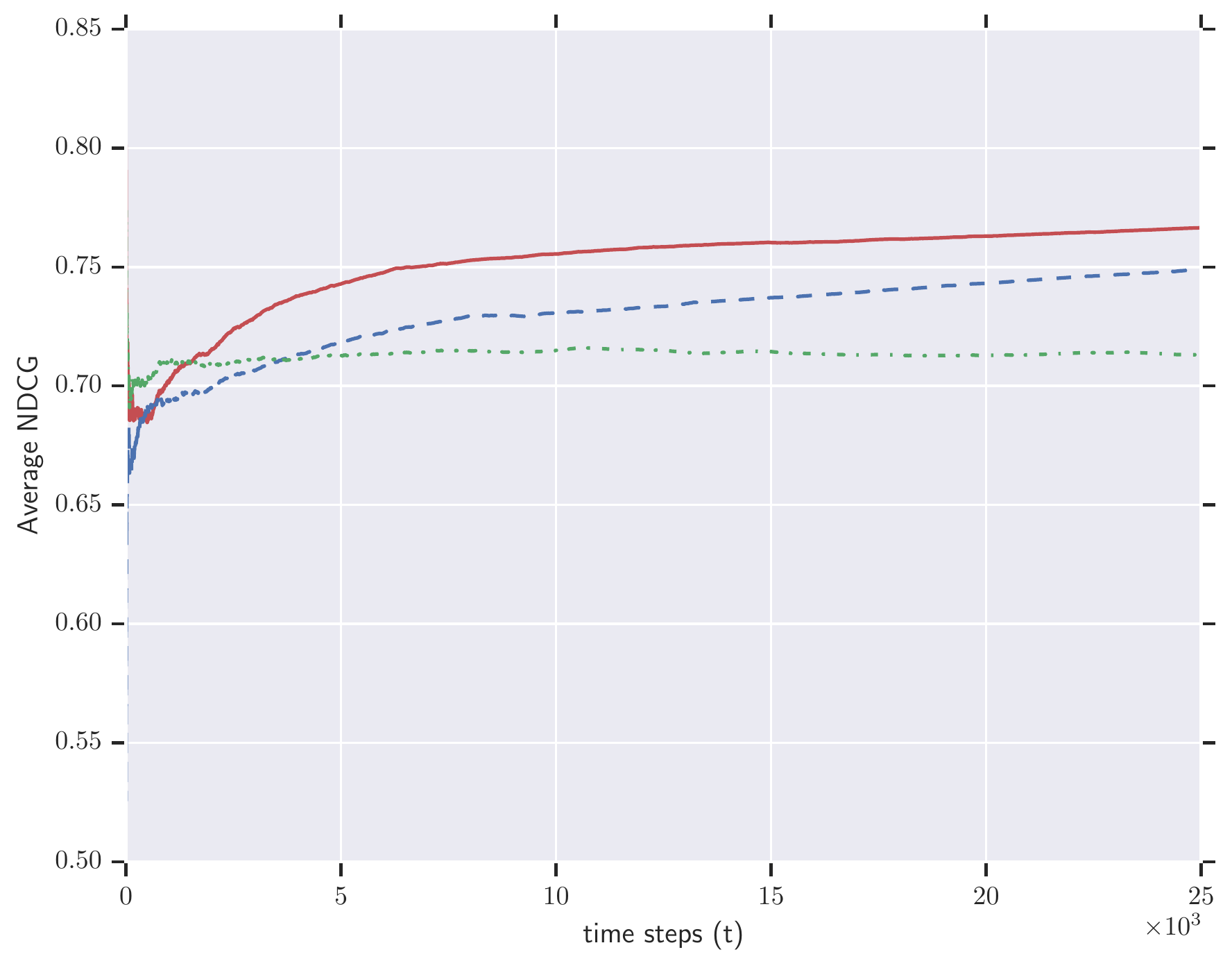} 
&
\includegraphics[width=1.9in]{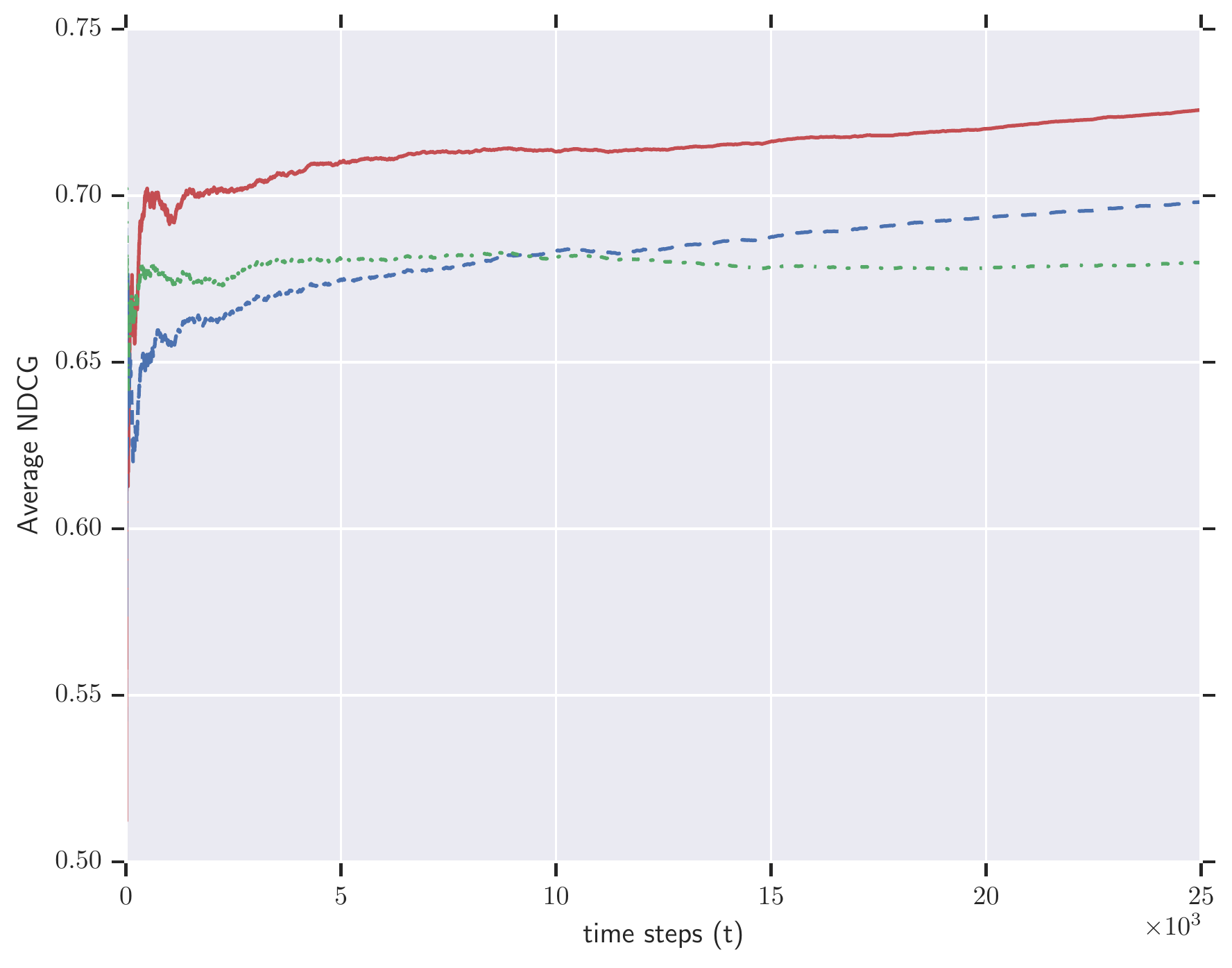} 
\end{tabular}
\caption{\small \label{fig:ndcg}
Evaluation of different algorithms in terms of average cumulative NDCG@5 for: Book Crossing (left), MovieLens (middle), and Jester (right) datasets.}
\end{figure*}

\textbf{Synthetic Datasets}. We consider three synthetic datasets: Gaussian, uniform, and Bernoulli. The Gaussian dataset is generated as follows. We first generate the user and item feature matrices $A$ and $B$ of rank $k = 5$ by drawing entries from Gaussian distributions $\mathcal{N}(0, \sigma_1^2)$ and $\mathcal{N}(0, \sigma_2^2)$, respectively. The true rating matrix is of size $200 \times 200$ and is obtained by $Y = A \cdot B^{\top}$. The observed ratings are then generated from $Y$ by adding a Gaussian noise $\mathcal{N}(0, \sigma^2)$ to the true ratings. In our experiments, $\sigma_1$, $\sigma_2$, and $\sigma$ for the Gaussian dataset are set to $1$, $1$, and $0.5$, respectively.

For both uniform and Bernoulli datasets, we first generate the true reward matrix $Y$ as follows. The user feature matrix $A$ is generated by picking each row uniformly at random from the $k$-dimensional simplex. Each entry of the item feature matrix $B$ is drawn uniformly at random from the interval $[0, 1]$. As with the Gaussian dataset, the true reward matrix is of size $200 \times 200$ and is equal to the product of the generated user and item feature matrices, i.e. $Y = A \cdot B^T$.
In the case of the uniform dataset, the observed ratings are drawn uniformly at random around the true reward values with a support of length $w$. In our experiments, we set $w = 0.5$. On the other hand, for the Bernoulli dataset, the observed ratings are Bernoulli samples with the given true reward values as the probability of success.

%For both the dataset as well as our algorithm, the parameters $\sigma_1$, $\sigma_2$, and $\sigma$ are set to $1$, $1$, and $0.5$, respectively. We also set the regularization parameters $\lambda = 0.1$. The optimal values of the regularization parameters depend on the size of the dataset; the larger is the dataset, a higher regularizer is required. 

We simulate the experiments according to the setting mentioned in the previous section; we assume that at time $t$, a random user $i_t$ arrives; the system recommends an item $j_t$ and the user rates the recommended item. The cumulative regret curves for different algorithms over different synthetic datasets are shown in Figure \ref{fig:synthetic}. 
As we can see from this figure, the proposed ALB algorithm outperforms PTS and NMF-Bandit by a large margin over all the three synthetic datasets. 
The tuned hyper-parameter pair $(\lambda, \sigma)$ for ALB over Gaussian, uniform, and Bernoulli datasets are $(0.01, 0.5)$, $(0.001, 0.5)$, $(1, 0.4)$, respectively. 
%Similarly, the hyper-parameter pair of number of particles and $\sigma$ for PTS over the three datasets are $()$, $()$, $()$, respectively. Finally, the tuned pair of $(\theta, m')$ for NMF-Bandit over the three datasets are $(2, 25)$, $(2, 20)$, $(3, 20)$, respectively.

%The experiments over synthetic datasets serves as a sanity check for the performance of the proposed ALB algorithm; from the latter experiments we can confirm that ALB produces a cumulative regret sub-linear in time.

%\begin{figure}[H]
%\center
%\includegraphics[width=3in, height=2in]{regret/gs_cumulative_regret.pdf} 
%\caption{\small \label{fig:gaussian}
%Cumulative Regret for the synthetic Gaussian dataset.}
%\end{figure}

\textbf{Real World Datasets}. Next, we perform experiments over three real world datasets: Book Crossing, MovieLens, and Jester. For Book Crossing, we pick a subset of $2000$ users and $2000$ books. In the case of MovieLens, we pick the $100K$ version of the dataset with $1000$ users and $1700$ movies. On the other hand, Jester has $150$ items (jokes) and we pick the first $5000$ users. 

In order to simulate an online recommender system with cold-start properties, we start from an empty set of ratings for all the users. We then iterate over the dataset in the following fashion: at any given time $t$, we assume that a random user $i_t$ arrives and the agent recommends an item $j_t$ constrained to the set of items rated by this user in the dataset. This experimental setting has been used in \cite{PTS}. Note that in order for NMF-Bandit to run in this setting, we need to restrict the randomized explorations to only include items with available ratings.

%The number of particles used for PTS is set to $30$ in all the experiments, as suggested in \cite{PTS}. 
%In addition, we use the same values suggested by \cite{PTS} for the PTS algorithm: $\sigma_1 = \sigma_2 = 1$, and $\sigma^2 = 0.5$.
%For the NMF-Bandit algorithm, the parameters $\theta$ (the exploration parameter) and $m'$ (size of the exploration set for each user, in their notation) are set to $3$ and $2 \times k$, respectively, as suggested in \cite{NMFBandit}. 
The tuned hyper-parameter pair of $(\lambda, \sigma)$ for our algorithm are set to $(0.1, 0.1)$, $(1, 0.4)$, and $(1, 0.9)$ for Book Crossing, MovieLens, and Jester, respectively. 
%This fixed set of values for the parameters $\lambda_1$, $\lambda_2$, and $\sigma$ have been sufficient to outperform both PTS and NMF-Bandit over different datasets. Further optimization of these parameters would only improve the results further.
%Figure \ref{fig:real} depicts the performance of the three algorithms for Book Crossing (left), MovieLens (middle), and Jester (right) datasets, which again indicate the superior performance of our algorithm in terms of cumulative regret. In particular, for the MovieLens dataset, we observe that the proposed algorithm outperforms the competing algorithms in terms of cumulative regret by a significant margin. We also note that PTS outperforms NMF-Bandit in almost all the experiments.
Figures \ref{fig:real} and \ref{fig:ndcg} depict the performance of the three algorithms for Book Crossing (left), MovieLens (middle), and Jester (right) datasets, which indicate the superior performance of our algorithm in terms of both cumulative regret and average cumulative NDCG. For the experiments on the Book Crossing dataset, the performance gaps in terms of NDCG between different algorithms are not as large as is the case for the other two datasets. This is to be expected since this dataset is particularly sparse, where many users have ratings available for a limited subset of books. For the other two datasets (MovieLens and Jester), we observe that ALB steadily outperforms the other algorithms by at least $0.03$ over time in terms of NDCG. In particular, for the MovieLens dataset, we observe that the proposed algorithm outperforms the competing algorithms in terms of cumulative regret by a significant margin. We also note that PTS outperforms NMF-Bandit in almost all the experiments.

\textbf{Choice of the Factorization Rank}. In the experiments presented so far, we assumed that the factorization rank $k$ is constant and is set to $5$. In this section, we evaluate the impact of the choice of the factorization rank on the performance of different algorithms. In doing so, we focus on the MovieLens dataset (as described in the previous section), and study the changes in the cumulative regret as we decrease or increase the factorization rank by $2$.

\begin{figure*}[t]
\center
\begin{tabular}{ccc}
\includegraphics[width=1.9in]{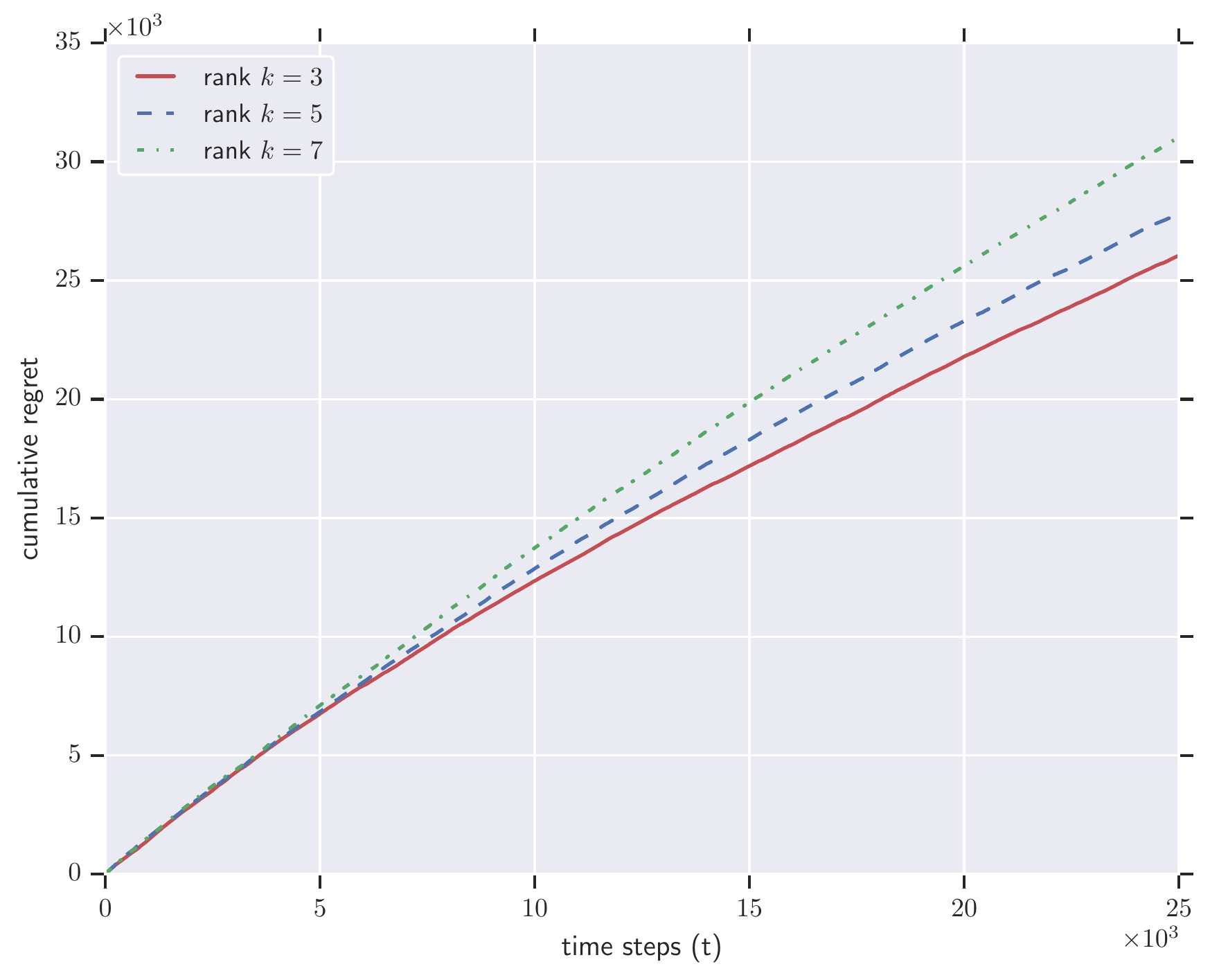} 
&
\includegraphics[width=1.9in]{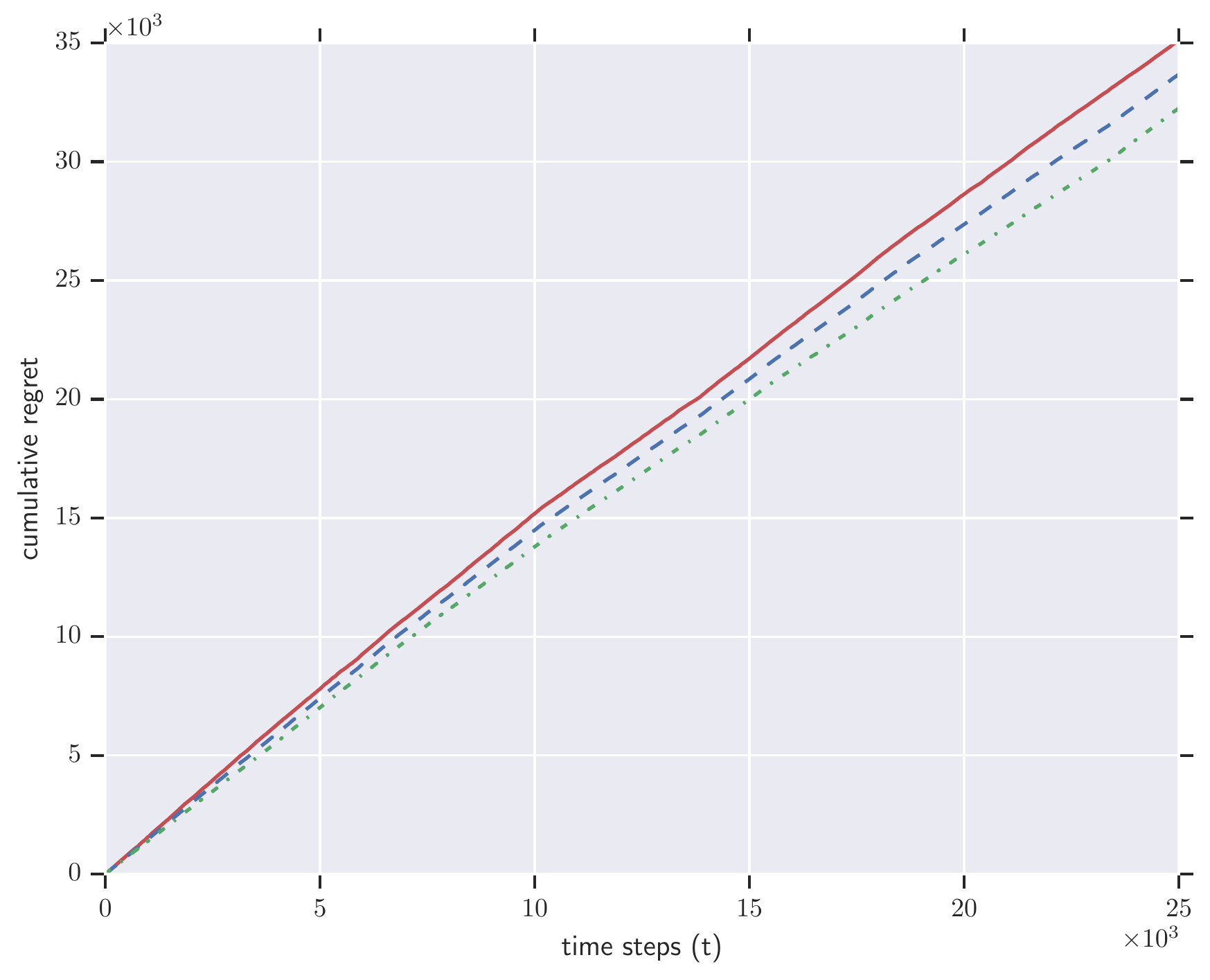} 
&
\includegraphics[width=1.9in]{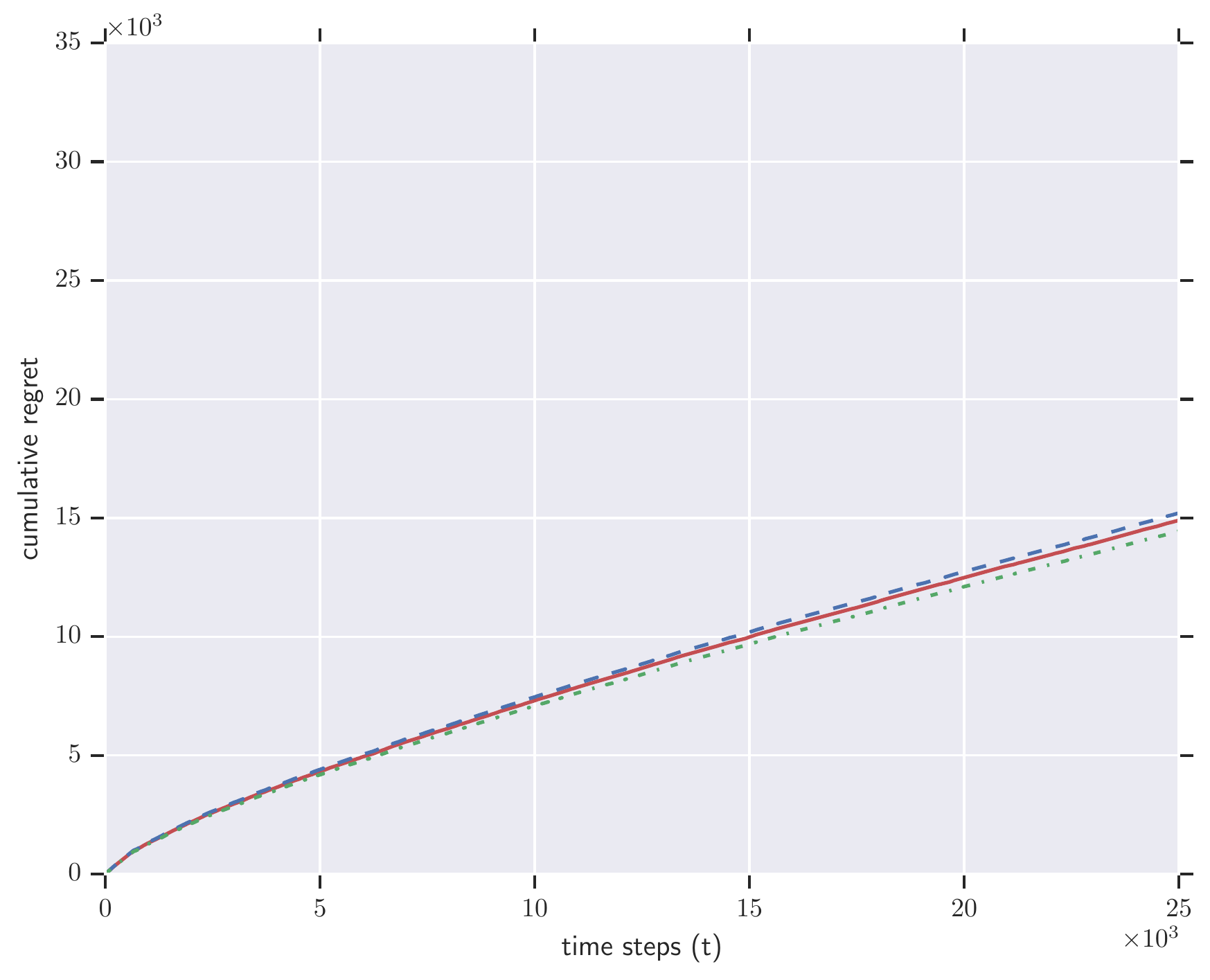} 
\end{tabular}
\caption{\small \label{fig:rank}
Evaluation of the impact of changes in the factorization rank on different algorithms in terms of cumulative regret: PTS (left), NMF-Bandit (middle), and ALB (right) algorithms.}
\end{figure*}

Figure \ref{fig:rank} depicts the performance of different algorithms in terms of cumulative regret over the set of factorization ranks $k \in \{3, 5, 7\}$ for PTS (left), NMF-Bandit (middle), and ALB (right) algorithms, respectively. As we can see from the figure, the performance of PTS improves slightly as we reduce the factorization rank. This is in contrast to the NMF-Bandit algorithm, whose performance improves slightly as we increase the factorization rank. Interestingly, we observe that the proposed ALB algorithm is more robust to the variations of the factorization rank, and only demonstrates minor changes in the cumulative regret as we reduce or increase the factorization rank $k$ over the given set. We note that regardless of the choice of the factorization rank from the test set $\{3, 5, 7\}$, the proposed ALB algorithm outperforms the competing methods in all the cases.

\section{Future Work}

Considering that the uncertainty in the reward distribution comes from not only the parameter vector (or user features vector) of the bandit, but also the action vector (or item features vector), it is not trivial to adapt
the existing regret guarantees in the linear bandit literature to the analysis of the proposed algorithm. As such, regret bound analysis for the proposed alternating linear bandit algorithm remains as a direction for future work. We note that an ideal algorithm for this problem should not only have a sub-linear dependence on time, but also have a sub-linear dependence on either the number of users or the number of items. PTS has such a regret guarantee but is only limited to the rank-$1$ case; even then, it has been assumed that the sampling approximation error is zero, which is not a realistic assumption. NMF-Bandit is the only algorithm with such a regret guarantee for the general rank $k$ case (which is yet not necessarily optimal), but in practice is outperformed by both PTS and our proposed ALB algorithm.

In addition, another direction for further work is to adopt the rarely switching version of the OFUL algorithm, rather than its basic form, to the proposed alternating linear bandits algorithm. The latter has the potential to reduce the computational complexity of the proposed algorithm significantly.

Finally, the focus of the numerical experiments in this paper has been on small to medium size datasets. We plan to perform further numerical experiments over different datasets of particularly larger size in the future.

% Bibliography
\bibliography{references}
\bibliographystyle{numeric}

\end{document}